# Reforming research funding:

# Combining editorial preregistration with grant peer review

Lutz Bornmann*, Gerald Schweiger$

* Science Policy and Strategy Department

Administrative Headquarters of the Max Planck Society

Hofgartenstr. 8,

80539 Munich, Germany.

Email: bornmann@gv.mpg.de

$ TU Wien

Karlsplatz 13,

1040 Wien

Email: gerald.schweiger@tuwien.ac.at

**Abstract**

Competitive grant funding is associated with high costs and a potential bias to favor conservative research. This comment proposes integrating editorial preregistration, in the form of registered reports, into grant peer review processes as a reform strategy. Linking funding decisions to in-principle accepted study protocols would reduce reviewer burden, strengthen methodological rigor, and provide an institutional foundation for (more) replication, theory-driven research, and high-risk research. Our proposal also minimizes strategic proposal writing and ensures scholarly output through the publication of preregistered protocols, regardless of funding outcomes. Possible implementation models include direct coupling of journal acceptance with funding, co-review mechanisms, voucher systems, and lotteries. While challenges remain in aligning journal and funding agency procedures, the integration of preregistration and funding offers a promising pathway toward a more transparent and efficient research ecosystem.

**Key words**

grant funding; peer review; preregistration; registered reports; research policy; funding reform; transparency

Modern research funding predominantly combines competitive project grants with institutional or block funding (Janger et al., 2019). In competitive schemes, investigators submit detailed proposals evaluated through peer review models—often independent expert assessments followed by panel deliberations—using broadly defined criteria such as "excellence" or "quality". The economic burden of competition is substantial: According to Graves et al. (2011), applicants bear roughly 85% of the total costs, while decision-making processes account for 10% and administrative functions for the remaining 5%. Reviewers and administrative staff also incur significant time and financial costs. Aczel et al. (2021) found that "the total time reviewers globally worked on peer reviews was over 100 million hours in 2020, equivalent to over 15 thousand years. The estimated monetary value of the time US-based reviewers spent on reviews was over 1.5 billion USD [United States Dollar] in 2020. For China-based reviewers, the estimate is over 600 million USD, and for UK-based, close to 400 million USD". Evidence suggests that high-risk, high-impact research is systematically under-funded in competitive environments, leading to more conservative research (Veugelers et al., in press).

Competitive grant systems also have ethical limitations. Studies have revealed low reliability and predictive validity in peer review, with top-ranked rejected proposals outperforming accepted proposals (in bibliometric measures) (Bornmann et al., 2010; Gallo & Glisson, 2018). Non-merit factors such as gender, nationality, and institutional affiliation can influence outcomes (Bornmann, 2011; Thorngate & Carroll, 1991), and the pressure to secure funding may encourage questionable research practices—ranging from salami-slicing as publication strategy to overstating proposed study confidence—while imposing mental health strains and career precarity on researchers, particularly early-career and underrepresented groups (Schweiger et al., 2024). The cumulative advantage effect, where previously successful applicants are more likely to receive future funding, entrenches inequalities and narrows the diversity of funded perspectives (Long & Fox, 1995; Thorngate et al., 2009).

These dynamics risk undermining the pluralism and exploratory freedom that is essential to robust scientific progress.

Beyond these concerns, competitive grant systems may disturb the epistemic landscape by privileging short-term deliverables over long-term inquiry. An emphasis on impact can marginalize replication studies, theoretical work, and interdisciplinary approaches that do not fit neatly into predefined (disciplinary) categories. Strategic behavior in proposal writing—such as tailoring research questions to perceived funder or disciplinary preferences (Stanford Research Development Office, 2025)—can lead to intellectual homogenization, slowed innovation processes, and reduction of knowledge gain.

Taken together, these weaknesses of competitive grant systems suggest the need for improvements that reorients funding decisions stronger toward methodological quality, efficiency, transparency, fairness, and risky research. Many proposals have been published to reform the grants peer review system (see, e.g., Forscher et al., 2019). In this comment, we would like to introduce a new avenue: the integration of funding decisions with journal-based study preregistration. Preregistrations are peer-reviewed and published registered reports of studies that are planned to be conducted. Our proposal addresses the shortcomings outlined above by shifting the evaluative focus from complex grant proposals to rigorously reviewed study protocols. Embedding funding decisions within the publication process may offer a mechanism to realign incentives and reduce distortions in competitive grant systems.

Registered reports are a publication format designed to enhance transparency and reduce bias in scientific research. The process typically involves two stages. In Stage 1, researchers submit a detailed study protocol—including the literature overview, research question, hypotheses, methodology, and planned analyses—which undergoes rigorous peer review before data collection begins. If the protocol is deemed methodologically sound, the journal issues an in-principle acceptance, committing to publish the final study regardless of the outcome, provided the authors adhere to the approved protocol. This approach has been established at several journals to prevent publication bias and p-hacking, and it encourages rigorous, theory-driven research.

For example, the journal *PLOS One* offers registered reports across multiple disciplines. Their model includes structured templates for protocols, transparent review criteria, and a commitment to publish null results. Registered reports differ from traditional preregistrations (e.g., via the Open Science Framework, OSF) in that they are embedded in a journal's publication pipeline and include formal peer review. While OSF preregistrations document study plans publicly, registered reports offer the additional benefit of editorial oversight and guaranteed publication upon successful completion. This makes them particularly suitable as a basis for funding decisions.

Upon successful peer review and in-principle acceptance by the journal (Stage 1), the study becomes eligible for funding by research agencies.

This approach directly mitigates several core weaknesses of the current system:

- **Reduction of reviewer burden**: The methodological evaluation is conducted once, by journal reviewers, reducing duplication and relieving pressure on grant peer review systems.
- **Improved epistemic quality**: Funding decisions are based on pre-approved study designs rather than speculative promises, promoting methodological rigor. Two common reasons for proposal rejection are "the proposal was not absolutely clear in describing one or more elements of the study … [and] the proposal was not absolutely complete in describing one or more elements of the study" (Locke et al., 2013, p. 188).
- **Support for replication and theory-driven research**: While preregistration frameworks can encourage replication and theory-driven studies by shifting attention toward methodological rigor rather than novelty or outcome, their actual funding will still depend on the priorities and policies of funding agencies.

If only a subset of preregistered protocols can be funded, replication studies may continue to face disadvantages unless agencies explicitly value and incentivize them. In this sense, our proposed integration does not automatically guarantee increased support for replication, but it provides an institutional structure that enhances the visibility of such studies. To realize this potential, funding agencies could complement the model by articulating clear strategic commitments to replication and foundational work—through dedicated funding streams, thematic programs, or explicit evaluation criteria—while leveraging preregistration to ensure methodological rigor and transparency.

- **Support for risky research**: Reviewers on grant panels tend to be risk-averse (Barlösius et al., 2023): High risk research may be seen as a waste of funds that justifies rejecting the application. As preregistration assessments focus primarily on the quality and rigor of study designs rather than financial considerations, reviewers are less constrained by funding concerns. Consequently, they may be more inclined to support novel or unconventional research approaches, provided the methodology is sound.
- **Minimization of bias and strategic behavior**: The focus shifts from persuasive grant writing to transparent methodological quality, reducing incentives for overstatement and tailoring to funder preferences.
- **Intrinsic publication value**: The act of preregistration itself constitutes a scholarly contribution, strengthening transparency and academic credit.
- **Guaranteed scholarly output**: Even if a study is not funded later on, the preregistered protocol remains a citable publication, ensuring visibility and recognition for the research idea.

Several implementation scenarios are conceivable:

1. **Direct coupling**: Funding agencies automatically support studies with Stage 1 acceptance at designated journals.
2. **Co-review model**: Journals and funders jointly evaluate protocols, with synchronized decisions.
3. **Voucher system**: Researchers receive funding vouchers upon journal acceptance, redeemable at participating agencies.
4. **Lottery system**: Funding is allocated through a lottery (Shaw, 2022) among preregistered and in-principle accepted study protocols that meet predefined quality standards. Empirical evidence suggests that while peer review can effectively filter out flawed proposals, it struggles to differentiate reliably among high-quality submissions (Shaw, 2022). A lottery mechanism could therefore promote fairness, mitigate bias, and enable a more transparent and efficient distribution of research funds.
5. **Community voting**: Funding decisions among preregistered and in-principle accepted study protocols are informed by democratic voting within the scientific community, where researchers collectively nominate peers whose work they consider most deserving of support (Barnett et al., 2017).

By integrating the evaluation of scientific study proposals with publication pathways, the integration of preregistration into grant peer review processes offers a possible alternative to the inefficient logic of many competitive grant peer review processes. The proposed integration may recalibrate incentives toward transparency, pluralism, and epistemic robustness, while reducing administrative overhead and enhancing the reliability of funding decisions. Even if applications are not funded, the ideas for a study will be published.

However, our approach also has limitations:

- It presupposes that the primary outcome of a research project is a publication, which may not apply to all types of research—particularly applications with infrastructural goals, studies with the goal of software development, applied research at advanced stages of technological readiness, or industrial research. The primary outputs in these projects may not be scientific publications but rather prototypes or technological innovations that form part of companies' intellectual property, with universities or research institutions often playing a supporting rather than a leading role.

- Large-scale projects that aim to produce multiple outputs may struggle to fit into a single preregistration framework. Conversely, this constraint could help reduce salami-slicing by encouraging researchers to consolidate their work into coherent, well-designed studies.
- The integration of preregistration into grant peer review processes and may require careful alignment with funding cycles. Since journal peer review timelines may not match the schedules of funding agencies, mechanisms for temporal coordination—such as rolling submissions, flexible funding windows, or conditional funding pending journal acceptance—would be necessary to ensure feasibility.
- Close coordination between journals and funding agencies is essential, which may be difficult to implement across heterogeneous institutional landscapes.

Despite these challenges, the integration of preregistration and funding may represent a promising step toward a more transparent, efficient, and epistemically sound research ecosystem with many advantages for funding agencies and applicants.

Future developments in this area targeting the described limitations may include the creation of shared digital platforms that link journal acceptance data with funding portals, enabling seamless verification and coordination between publication and funding processes. Such platforms would allow funding agencies to automatically recognize in-principle accepted study protocols and streamline the allocation of resources.

Standardized metadata for preregistered protocols could be developed to ensure interoperability across journals and funders. This would facilitate automated tracking, evaluation, and integration of study plans into funding workflows, reducing administrative burden and enhancing transparency.

Pilot programs could serve as testbeds for the integration of preregistration into grant peer review processes. By implementing coordinated review and funding mechanisms in selected fields, institutions could assess feasibility, identify procedural bottlenecks, and refine interfaces between journals and funding bodies. Over time, these pilots may evolve into scalable models for a more integrated and epistemically robust research infrastructure.